\documentclass[runningheads]{llncs}
\usepackage{graphicx}
\usepackage{tikz}
\usepackage{comment} 
\usepackage{amsmath,amssymb} % define this before the line numbering.
\usepackage{color}
\usepackage{graphicx}
\usepackage{epsfig}
\usepackage{mathtools}
\usepackage{iac_pkg}
\usepackage{multirow}
\usepackage{tabularx, booktabs}
\usepackage{subfig}
\usepackage{color}
\usepackage{wrapfig}
\usepackage{xspace}
\usepackage{threeparttable}
\usepackage{float}

\usepackage[width=120mm,left=12mm,paperwidth=146mm,height=193mm,top=12mm,paperheight=217mm]{geometry}

\definecolor{citecolor}{RGB}{34,139,34}
\usepackage[pagebackref=true,breaklinks=false,citecolor=citecolor,colorlinks,bookmarks=false]{hyperref}
\usepackage[capitalise]{cleveref}

\begin{document}
% \linenumbers
\pagestyle{headings}
\mainmatter

\title{{\normalsize Rapid quantification of COVID-19 pneumonia burden from computed tomography with convolutional LSTM networks.}}

\titlerunning{Rapid quantification of COVID-19 pneumonia burden}
\author{Kajetan Grodecki\inst{*1} \and
Aditya Killekar\inst{*2} \and Andrew Lin\inst{1} \and Sebastien Cadet\inst{2} \and Priscilla McElhinney\inst{1} \and Aryabod Razipour\inst{1} \and Cato Chan\inst{3} \and Barry D. Pressman\inst{3} \and Peter Julien\inst{3} \and Judit Simon\inst{4} \and Pal Maurovich-Horvat\inst{4} \and Nicola Gaibazzi\inst{5} \and Udit Thakur\inst{6} \and Elisabetta Mancini\inst{7} \and Cecilia Agalbato\inst{7} \and Jiro Munechika\inst{8} \and Hidenari Matsumoto\inst{9} \and Roberto Men\`e\inst{10,11} \and Gianfranco Parati\inst{10,11} \and Franco Cernigliaro\inst{10,11} \and Nitesh Nerlekar\inst{6} \and Camilla Torlasco\inst{10,11} \and Gianluca Pontone\inst{7} \and Damini Dey\inst{1} \and Piotr J. Slomka\inst{2}
}

\authorrunning{K. Grodecki et al.}

\institute{Biomedical Imaging Research Institute, Cedars-Sinai Medical Center, Los Angeles, CA, USA  \and Department of Medicine, Cedars-Sinai Medical Center, Los Angeles, CA, USA \and Department of Imaging, Cedars-Sinai Medical Center, USA \and Department of Radiology, Semmelweis University, Budapest, Hungary \and Cardiology, Azienda Ospedaliero-Universitaria di Parma, Parma, Italy \and Monash Health, Melbourne, Australia \and Centro Cardiologico Monzino IRCCS, University of Milan, Italy \and Division of Radiology, Showa University School of Medicine, Tokyo, Japan \and Division of Cardiology, Showa University School of Medicine, Tokyo, Japan \and Department of Cardiovascular, Neural and Metabolic Sciences, IRCCS Istituto Auxologico Italiano, Milan, Italy \and Department of Medicine and Surgery, University of Milano-Bicocca, Italy}

%\end{comment}
%******************
%{\let\thefootnote\relax\footnotetext{\textsuperscript{*} indicates equal contribution}}

\maketitle
\nnfootnote{\textsuperscript{*} indicates equal contribution}
\begin{flushleft}
\vspace{5em}
Corresponding author:
\newline Piotr J. Slomka
\newline Division of Artificial Intelligence in Medicine
\newline Department of Medicine
\newline Cedars-Sinai Medical Center 
\newline 8700 Beverly Blvd, Los Angeles, CA 90048
\newline Email: \href{mailto:Piotr.Slomka@cshs.org}{Piotr.Slomka@cshs.org}
\newline Tel: +1 310 423 4348 
\newline Fax: +1 310 423 0173
\end{flushleft}
\clearpage
%%%%%%%%% ABSTRACT
\begin{abstract}
Quantitative lung measures derived from computed tomography (CT) have been demonstrated to improve prognostication in Coronavirus disease 2019 (COVID-19) patients, but are not part of the clinical routine since required manual segmentation of lung lesions is prohibitively time-consuming. We propose a new fully automated deep learning framework for quantification and differentiation between lung lesions in COVID-19 pneumonia from both contrast and non-contrast CT images using convolutional Long Short-Term Memory (LSTM) networks. Utilizing the expert annotations, model training was performed using 5-fold cross-validation to segment ground-glass opacity and high opacity (including consolidation and pleural effusion). The performance of the method was evaluated on CT data sets from 197 patients with positive reverse transcription polymerase chain reaction test result for SARS-CoV-2.  Strong agreement between expert manual and automatic segmentation was obtained for lung lesions with a Dice score coefficient of 0.876 \textpm\xspace 0.005; excellent correlations of 0.978 and 0.981 for ground-glass opacity and high opacity volumes. In the external validation set of 67 patients, there was dice score coefficient of 0.767 \textpm\xspace 0.009 as well as excellent correlations of 0.989 and 0.996 for ground-glass opacity and high opacity volumes. Computations for a CT scan comprising 120 slices were performed under 2 seconds on a personal computer equipped with NVIDIA Titan RTX graphics processing unit. Therefore, our deep learning-based method allows rapid fully-automated quantitative measurement of pneumonia burden from CT and may generate the big data with an accuracy similar to the expert readers. 
\end{abstract}
\section{Introduction}\label{sec:intro}
Coronavirus disease 2019 (COVID-19) is a global pandemic and public health crisis of catastrophic proportions, with over 123 million confirmed cases worldwide as of March 22  2021 ~\cite{who}. While vaccines are now available–they are not 100\% effective, new strains are emerging and not all the population will be vaccinated. It is likely that annual vaccinations will be necessary and continuous monitoring for the disease will be needed.  While the diagnosis of COVID-19 relies on a reverse transcription polymerase chain reaction (RT-PCR) test in respiratory tract specimens, computed tomography (CT) remains the central modality in disease staging~\cite{boger2020systematic,francone2020chest,khatami2020meta}. Specific CT lung features include peripheral and bilateral ground-glass opacities (GGO), with round and other specific morphology as well as peripheral consolidations and increasing extension of such opacities has been associated with the risk of critical illness\cite{wang2020temporal,pan2020time,bernheim2020chest}. Whereas conventional visual scoring of the COVID-19 pneumonia extent correlates with clinical disease severity, it requires proficiency in cardiothoracic imaging and ignores lesion features such as volumes, density, or inhomogeneity\cite{yang2020chest,li2020ct}. On the other hand, CT-derived quantitative lung measures are not part of the clinical routine, despite being demonstrated to improve prognostication in COVID-19 patients, due to prohibitively time-consuming manual segmentation of the lung lesions required for computation \cite{grodecki2020quantitative,gieraerts2020prognostic,chaganti2020automated}. 

Deep learning, a class of artificial intelligence (AI), has shown to be very effective for automated object detection and image classification from a wide range of data\cite{commandeur2018deep}. A variety of AI systems has been introduced to aid radiologists in the detection of lung involvement in COVID-19 with several presenting the potential to improve the performance of junior radiologists to the senior level \cite{gieraerts2020prognostic,zhang2020clinically}. However, most medical segmentation networks consume a lot of memory in storing the intermediate features for skip connections\cite{cciccek20163d,milletari2016v}. Considering the spatiotemporal nature of CT images, utilization of adjacent slices of input may improve performance in semantic segmentation tasks. Convolutional long short-term memory (ConvLSTM) networks have the capability of preserving relevant features with simultaneous dismission of irrelevant ones in the form of the feedback loop, which translates into a memory-sparing strategy for the holistic analysis of the images. In this paper, we employ the ConvLSTM to facilitate rapid segmentation and accurate 3D quantification of the disease involvement of lung lesions in COVID-19 pneumonia from both contrast and non-contrast CT images. 

\section{Methods}\label{sec:methods}
\subsection{Patient population}\label{sec:pat_pop}
The cohort used in this study comprised 264 patients, who underwent chest CT and had a positive RT-PCR test result for SARS-CoV-2. A total of 197 patients were included into training cohort and 67 were used for external validation. Datasets for 187 out of 197 patients from the training cohort were collected from the prospective, international, multicenter registry involving centers from North America (Cedars-Sinai Medical Center, Los Angeles, USA [n=75]), Europe (Centro Cardiologico Monzino [n = 64], and Istituto Auxologico Italiano [n = 17]; both Milan, Italy), Australia (Monash Medical Centre, Victoria, Australia [n=6]) and Asia (Showa Medical University, Tokyo, Japan [n=25]), where either non-contrast [n=157] or contrast-enhanced [n=30] chest CT was performed to aid in the triage of patients with a high clinical suspicion for COVID-19, in the setting of a pending or comorbidities associated with severe illness from COVID-19. The population is described in Table 1. Datasets for the remaining 10 COVID-19 patients were derived from an open-access repository of non-contrast CT images, therefore no clinical data were provided for this cohort. Out of 31,560 transverse slices available 15,588 had lesions \cref{table:table2}. External validation cohort comprised 67 non-contrast CT scans of COVID-19 patients: 50 from an open-access repository \cite{morozov2020mosmeddata} and 17 additional from Italy (Centro Cardiologico Monzino). There were 12,102 transverse slices available in this cohort and 6,503 had lesions \cref{table:table2}.  All data were deidentified prior to being enrolled in this study.   The CT images from each patient and the clinical database were fully anonymized and transferred to one coordinating center for core lab analysis. The study was conducted with the approval of local institutional review boards (Cedars-Sinai Medical Center IRB\# study 617) and written informed consent was waived for fully anonymized data analysis.
\subsection{Imaging Protocol}\label{sec:imag_proto}
Chest CT scans were performed with different multi-slice CT systems: Aquilion ONE (Toshiba Medical Systems, Otawara, Japan); GE Revolution, GE Discovery CT750 HD, or LightSpeed VCT (GE Healthcare, Milwaukee, WI, USA); and Brilliance iCT (Philips Healthcare, Cleveland, OH, USA). Parameters used for scans without intravenous contrast included a peak x-ray tube voltage of 120 kV, automatic tube current modulation (300-500 mAs), and slice thickness of 0.625 to 1.25 mm. The protocol for contrast-enhanced included a peak x-ray tube voltage of 120 kV, automatic tube current modulation (500-650 mAs), and slice thickness of 0.625 to 1.0 mm. A total of 80–100 ml iodinated contrast material (Iomeron 400, Bracco Imaging SpA, Milan, Italy; or Ominpaque 350, GE Healthcare, United States) was injected intravenously at a rate of 5 ml/s and followed by 20-30ml of saline chaser at a rate of 4-5 ml/s. Images were reconstructed using standard lung filters specific to each CT vendor. All scans were obtained in the supine position during inspiratory breath-hold.
\subsection{Ground truth generation}\label{sec:gt_gen}
Images were analyzed at the Cedars-Sinai Medical Center core laboratory by two physicians (K.G. and A.L.) with 3 and 8 years of experience in chest CT, respectively, and who were blinded to clinical data. A standard lung window (width of 1500 Hounsfield units [HU] and level of $-400$ HU) was used. Lung abnormalities were segmented using semi-automated research software (FusionQuant Lung v1.0, Cedars-Sinai Medical Center, Los Angeles, CA, USA). These included ground glass opacities (GGO), consolidation, or pleural effusion according to the Fleischner Society lexicon. Consolidation and pleural effusion were collectively segmented as high-opacity to facilitate the training of the model due to a limited number of slices involving these lesions. Chronic lung abnormalities such as emphysema or fibrosis were excluded from segmentation, based on correlation with previous imaging and/or a consensus reading. GGO was defined as hazy opacities that did not obscure the underlying bronchial structures or pulmonary vessels; consolidation as opacification obscuring the underlying bronchial structures or pulmonary vessels; and pleural effusion as a fluid collection in the pleural cavity. Total pneumonia volume was calculated by summing the volumes of the GGO and consolidation components. Total pneumonia burden was calculated as: total pneumonia volume/total lung volume×100\%. Difficult cases of quantitative analysis were resolved by consensus.
\subsection{Data preparation for convolutional neural networks}\label{sec:data_prep}
\begin{figure}[ht!]
    \caption{{\bf Framework for data preparation.}  First, input slices were restricted to the body region and then uniformly resized to 256x256. Further, the data was randomly augmented with the rotation of $[-10, +10]$ degrees, translation of up to 10 pixels in x- and y-directions, and scaling of [0.9, 1.05] times. Finally, data were homogenized by clipping the Hounsfield units between $-1024$ to 600 (Lung region) and normalizing between 0 and 1 using Min-Max scaling. 
    }
    \vspace{-4em}
    \begin{center}
    \includegraphics[width=\linewidth]{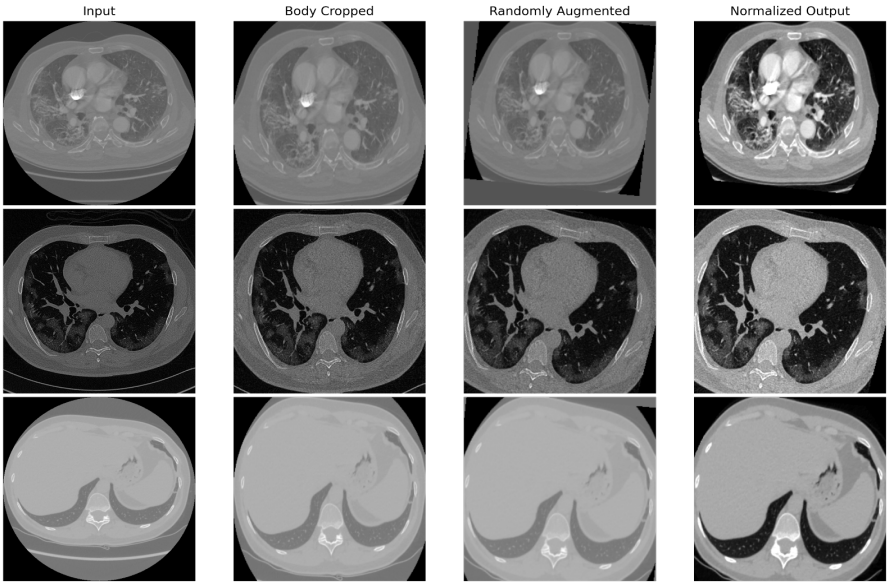}
    \end{center}
    \label{fig:preprocessing}
\end{figure}

First, the input slices were cropped to the body region and then uniformly resized to 256x256 in order to optimize the computation cost and performance \cref{fig:preprocessing}. Further, the data was randomly augmented with the rotation of $[-10, +10]$ degrees, translation of up to 10 pixels in x- and y-directions, and scaling of $[0.9, 1.05]$ times. Finally, the data were homogenized by clipping the Hounsfield units between $-1024$ to 600 (Lung region) and normalizing between 0 and 1 using Min-Max scaling.
\subsection{Model architecture}\label{sec:arch}
\begin{figure}[ht!]
    \caption{{\bf Framework of the proposed method.} 
    }
    \vspace{-2em}
    \begin{center}
        \includegraphics[width=\linewidth]{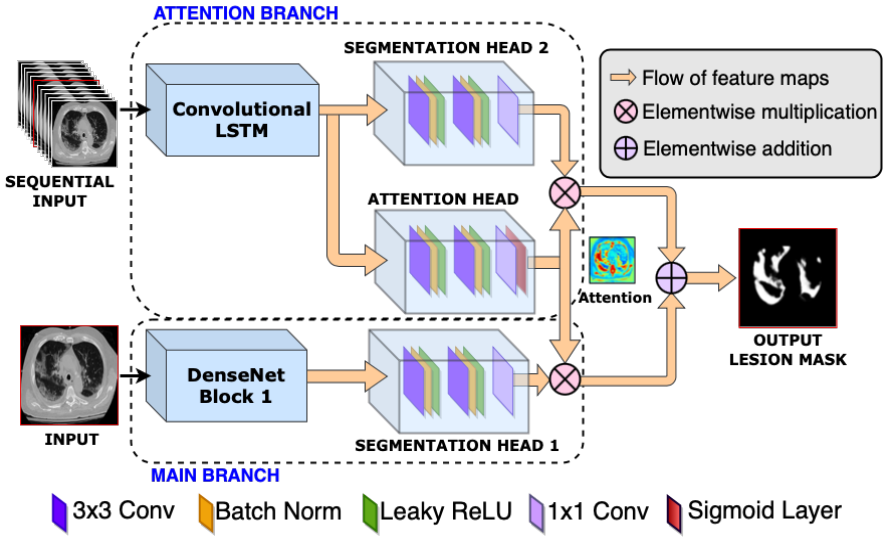}
    \end{center}
    \label{fig:arch}
\end{figure}

The model architecture was based on hierarchical multi-scale attention for semantic segmentation with the attention head looking through several adjacent slices above and below the current one to improve lesion recognition \cite{tao2020hierarchical}. As shown in \cref{fig:arch}, the architecture was divided into two branches: the main segmentation branch and the attention branch. The main segmentation branch, consisting of a dense block followed by a segmentation head, extracted larger and easy-to-classify lesions\cite{huang2017densely}. The attention branch comprised a sequential processor: a ConvLSTM followed by a segmentation head and an attention head\cite{shi2015convolutional}. The ConvLSTM block allowed to imitate a radiologist reviewing adjacent slices of a CT scan to detect lung abnormalities and ensure appropriate annotation.  The attention head synergized with the ConvLSTM to correct borderline misclassifications of the main segmentation branch. The segmentation head comprised three blocks: the first two blocks consisted of a 3x3 convolutional layer followed by a batch normalization layer and a LeakyRelu activation layer, while the final block was a 1x1 convolutional layer\cite{xu2015empirical}. The attention head was identical to the segmentation head in structure with the only difference being followed by an additional Sigmoid layer. A single slice intended for the segmentation was used as an input by the main segmentation branch, whereas the attention branch incorporated additional 5 slices above and below the target one. Thus, the ConvLSTM block processed sequentially a total of 11 slices and conveyed the final output to the attention and segmentation head of the attention branch. The output of the attention head was then multiplied with the respective semantic feature maps of the two segmentation heads.
\begin{equation}
    \setlength\abovedisplayskip{5pt}
    \setlength\belowdisplayskip{5pt}
    \mathbf{S}_\mathrm{out} = \mathbf{S}_\mathrm{main}\alpha + \mathbf{S}_\mathrm{attn}(1 - \alpha)
    \label{eq:merge}
\end{equation}
\subsection{Network Training}\label{sec:training}
The entire model was initialized using Kaiming He initialization, except for the dense block, where the pre-trained weights from ImageNet were used \cite{he2015delving,krizhevsky2012imagenet}. Region Mutual Information (RMI) loss was used as the cost function\cite{zhao2019region}. The model parameters were optimized using a Stochastic Gradient Descent optimizer with a momentum of 0.9, the initial learning rate of $10^{-3}$, and weight decay of $10^{-6}$. The learning rate was gradually decreased using Reduce on Plateau technique with a factor of 0.1 and patience of 10. This learning rate scheduler kept track of validation loss. In cases, where no improvement was seen for a “patience” number of epochs, the learning rate was reduced by the given “factor”. The training batch size was 32. The training was stopped as soon as the learning rate reached $10^{-7}$.
\subsection{Cross-validation (repeated testing)}\label{sec:cv}
\begin{figure}[ht!]
    \caption{{\bf Cross-Validation} For each fold of the 5-fold cross-validation, the following datasets were used: (1) training dataset (125 or 126 cases); (2) validation dataset (32 cases); (3) test dataset (39 or 40 cases).  
    }
    \vspace{-2em}
    \begin{center}
        \includegraphics[width=\linewidth]{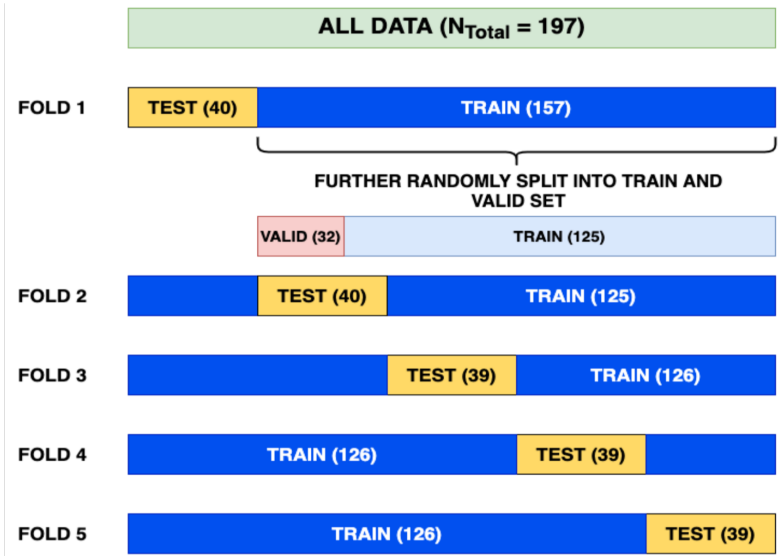}
    \end{center}
    \label{fig:cv}
\end{figure}

\sloppy To perform a robust non-biased evaluation of the framework, 5-fold cross-validation was used, using 5 separate models and 5 exclusive hold-out sets each of 20\%. The whole cohort of 197 cases was split into 5 subsets \cref{fig:cv}. For each fold of the 5-fold cross-validation, the following datasets were used: (1) training dataset (125 or 126 cases) was used to train the ConvLSTM; (2) validation dataset (32 cases) was defined to tune the network, select optimal hyperparameters and verify there was no over-fitting; (3) test dataset (39 or 40 cases) was used for the evaluation of the method. The final results were then concatenated from 5 separated subsets. Thus, the overall test population was 197 with 5 different models.
\subsection{Evaluation and statistical analysis}\label{sec:eval}
The primary end-point of this study was the performance of the deep learning method compared to the evaluation by the expert reader. Statistical analysis was performed using the SPSS software, version 23 (IBM SPSS Statistics, New York, USA). Continuous variables were expressed as mean ± standard deviation or median and interquartile range (IQR). Deep learning and expert quantifications were systematically compared using the Spearman correlation coefficient, Bland-Altman plots, and paired Wilcoxon rank-sum test. A p-value of $<$ 0.05 was considered statistically significant. Deep learning performance was also evaluated with the Dice similarity coefficient (DSC), as a measure of overlap between expert and deep learning. This is computed by $DSC = 2TP/(2TP+FP+FN)$, where TP (true positive) is the number of correctly positively classified voxels, FP (false positive) is the number of incorrectly positively classified, and FN (false negative) is the number of incorrectly negatively annotated classified voxels.
\subsection{Implementation}\label{sec:impl}
The code was written in Pytorch (v1.7.1) deep learning framework and incorporated in research CT lung analysis software (Deep Lung) written in C++. The training was performed on Nvidia Titan RTX 24GB GPU on a 10\textsuperscript{th} generation Intel Core i9 CPU. Deep Lung can be used with or without the  GPU acceleration.

\section{Results}\label{sec:results}
\subsection{Lesion quantification in the training cohort}\label{sec:train_cohort}
\begin{figure}[ht!]
    \caption{{\bf Comparison of expert and automatic lung lesion segmentation.} Blue represents ground-glass opacity. The dice score coefficient for this patient was 0.857. 
    }
    \vspace{-6em}
    \begin{center}
        \includegraphics[width=\linewidth]{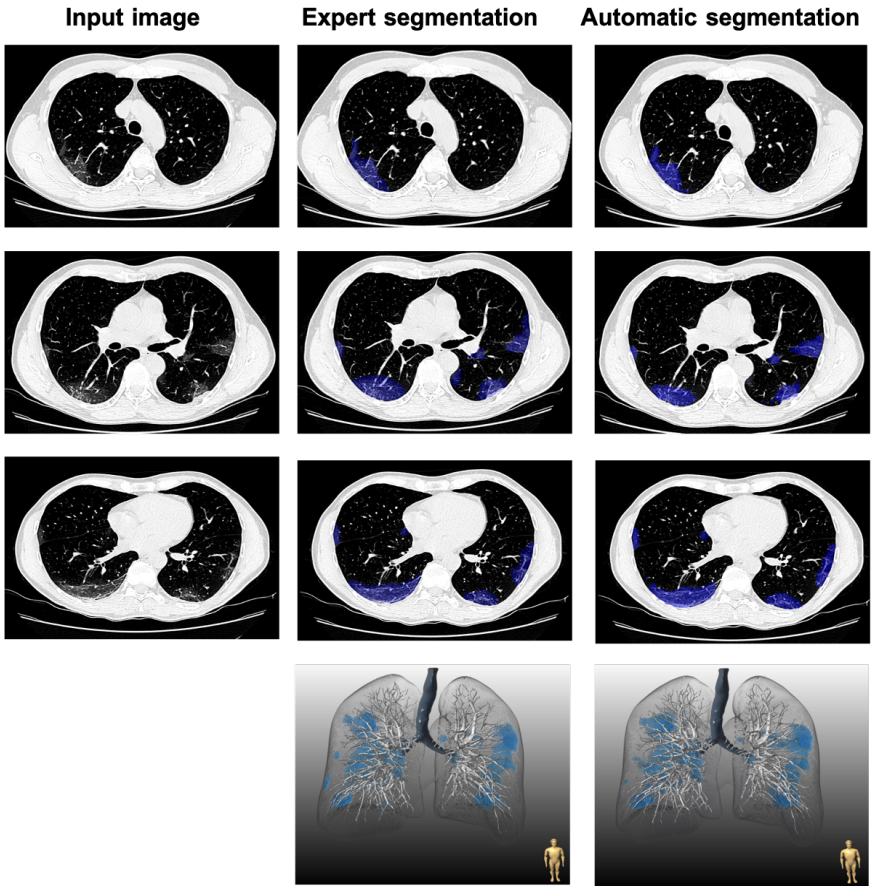}
    \end{center}
    \label{fig:comp_1}
\end{figure}

\begin{figure}[ht!]
    \caption{{\bf Comparison of expert and automatic lung lesion segmentation.} Blue represents ground-glass opacity and yellow represents consolidations. The dice score coefficient for this patient was 0.792. 
    }
    \vspace{-8em}
    \begin{center}
        \includegraphics[width=\linewidth]{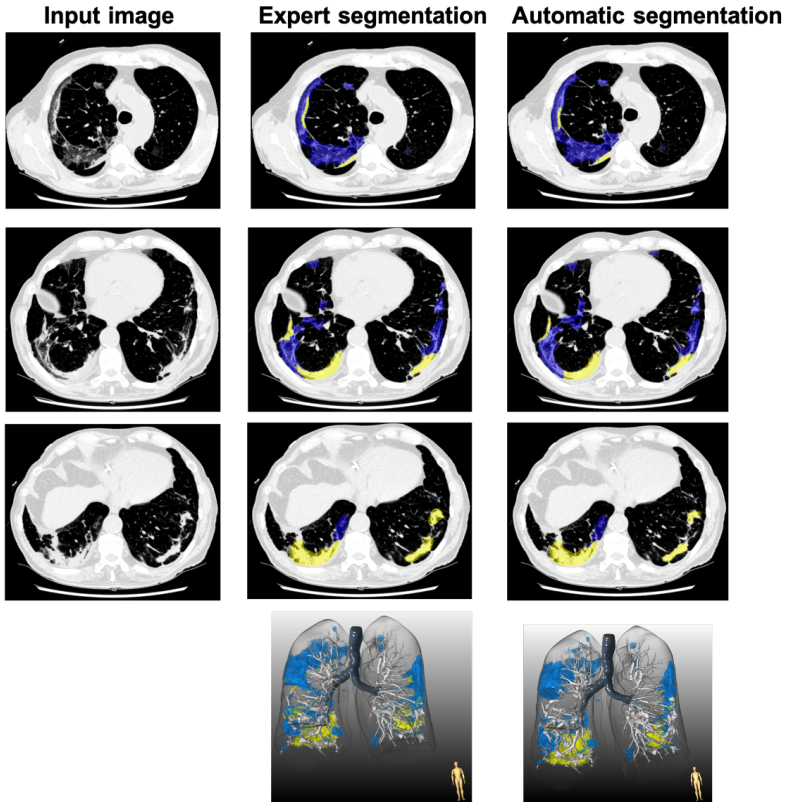}
    \end{center}
    \label{fig:comp_2}
\end{figure}

\begin{figure}[ht!]
    \caption{Bland-Altman plots and Spearman correlation for volumes of ground-glass opacity (A-B) and high-opacity  (C-D) between expert and automatic quantification in a testing cohort. 
    }
    \vspace{-2em}
    \begin{center}
        \includegraphics[width=\linewidth]{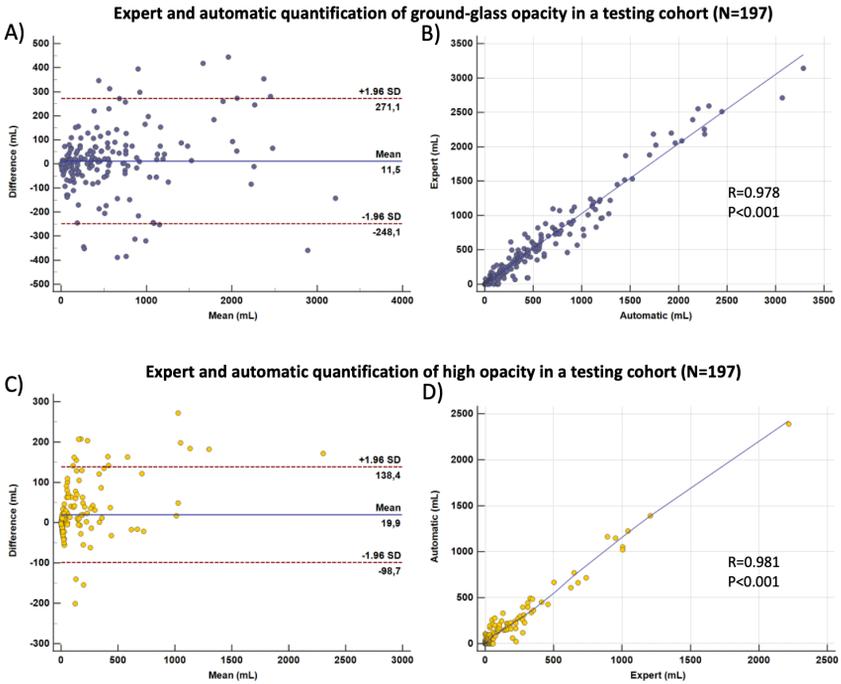}
    \end{center}
    \label{fig:BA_plot}
\end{figure}

A comparison between expert and automatic segmentation of the lung and its lesions is presented in \cref{fig:comp_1,fig:comp_2}. Across the 5-fold test dataset, a mean DSC of 0.876 \textpm\xspace 0.005 was obtained for the lesion mask differentiating between GGO and high-opacities. The median GGO volume was 330.1 mL (IQR 85.8 – 730.5 mL) and 326.1 mL (IQR 81.7 – 731.5 mL; p=0.849) for expert and automatic segmentations, respectively. An excellent correlation of 0.978 was obtained (p$<$0.001) and the Bland-Altman analysis demonstrated a low bias of 11.5mL \cref{fig:BA_plot}(A-B). Similarly, no significant differences between expert and automatic segmentation were found for high-opacities volumes (6.4 mL [IQR 0 – 151.1 mL] vs 12.1 [IQR 0.3 – 88.1 mL]; p=0.658); with excellent correlation (r=0.981, p$<$0.001) and a non-significant bias of 19.9 mL \cref{fig:BA_plot}(C-D).
\subsection{Lesion quantification in the external validation cohort}\label{sec:val_cohort}
A mean DSC of 0.767 \textpm\xspace 0.009 was obtained for the lesion mask differentiating between GGO and high-opacities in the external validation dataset using the model with the highest mean DSC. The median GGO volume was 74.9 mL (IQR 25.8 – 147.52 mL) and 90.2 mL (IQR 30.2 – 175.2 mL; p=0.849) for expert and automatic segmentations, respectively. An excellent correlation of 0.989 was obtained (p$<$0.001) and the Bland-Altman analysis demonstrated a low bias of 6.8mL (Supplementary \cref{fig:supp_fig1} A-B). Similarly, no significant differences between expert and automatic segmentation were found for high-opacities volumes (0 mL [IQR 0 – 0 mL] vs 0 mL [IQR 0.0 – 1.3 mL]; p=0.658); with excellent correlation (r = 0.996, p$<$0.001) and non-significant bias of 0.2 mL (Supplementary \cref{fig:supp_fig1} C-D).
\subsection{Lesion quantification in non-contrast and contrast-enhanced CT}\label{sec:enhanced_ct_result}
There were no significant differences between mean DSC calculated for segmentation from non-contrast (0.886 \textpm\xspace0.034) and contrast-enhanced (0.829 \textpm\xspace 0.029; p$<$0.001) CT scans. The median difference between expert and automatic measurements of GGO (14.7 mL [IQR $-34.2$ – 23.3 mL] vs $-10.3$ mL [IQR $-62.3$ – 16.4 mL]; p=0.841) and high opacities ($-11.3$ mL [IQR $-48.7$ – 18.3 mL] vs 0.0 mL [IQR $-21.3$ – 0.5 mL]; p=0.667) were comparable between scanning protocols.
\subsection{Computational performance}\label{sec:performance}
Minimal system requirements for the DeepLung are 8GB of RAM and 4GB of GRAM, thus allowing to run the segmentation in the lower-end hardware \cref{table:table3}. Computations for a CT scan comprising 120 slices were performed under 2 seconds on a personal computer equipped with NVIDIA Titan RTX (24GB DDR6) GPU and under 24 seconds using the unit with Nvidia Quadro K1200 (4GB DDR5) GPU. Time required for the computations without GPU acceleration was 150 seconds using Intel Core i7-6800K (3.4GHz) Manual segmentation, depending on the case complexity, required 15-20 minutes.
\subsection{Model Comparison}\label{sec:model_comparison}
In \cref{table:table4}, we show the performance of our model as compared to Unet2d and Unet3d. The performance is measured with two main metrics: mean dice score and compute resource utilization. The dice score is the cumulative mean dice score over test sets of 5-fold cross-validation (\cref{fig:cv}). The compute time and memory are calculated over 16 CT slices.
In this comparison, we see that our model slightly outperforms Unet2d and Unet3d in High-opacities and has comparable performance in segmenting ground-glass opacities. We also see that our model consumes significantly less memory and is slightly faster on GPU and much faster on CPU, than other models in comparison.
\section{Discussion}\label{sec:discussion}
We developed and evaluated a novel deep-learning ConvLSTM network approach for fully automatic quantification of COVID-19 pneumonia burden from both non-contrast and contrast-enhanced chest CT. To the best of our knowledge, ConvLSTM has not been applied before for segmentation of medical imaging data. We demonstrated that automatic pneumonia burden quantification by the proposed method shows strong agreement with expert manual measurements and rapid performance, suitable for clinical deployment. While vaccines have been developed to protect from COVID-19, the incidental findings of COVID-19 abnormalities due to imperfect vaccination rates and new strains will be a mainstay of medical practice. This method will provide a ‘real-time’ COVID-19 disease detection to the physician and allow image-guidance of novel therapies in severe cases.

The evolution of deep learning applications for COVID-19 is reflecting the changing role of CT imaging during the pandemic. Initially, when RT-PCR testing was unavailable or delayed, chest CT was used as a surrogate tool to identify suspected COVID-19 cases \cite{ai2020correlation}. AI-assisted image analysis could improve the diagnostic accuracy of junior doctors in differentiating COVID-19 from other chest diseases including community-acquired pneumonia and facilitate prompt isolation of patients with suspected SARS-CoV-2 infection \cite{zhang2020clinically,li2020using}. 

Currently, when RT-PCR testing is widely available with timely results, rapid quantification of pneumonia burden from chest CT as proposed here can be used for the management of the patients with COVID-19. As demonstrated in prior investigations, increasing attenuation of GGO and a higher proportion of consolidation in total pneumonia burden had prognostic value, thus underscoring the importance of utilizing all CT information for training the patients \cite{grodecki2020quantitative,grodecki2021epicardial}. Manual segmentation of the lung lesions is, however, challenging and prohibitively time-consuming task due to complex appearances and ambiguous boundaries of the opacities \cite{wang2020noise}. To automate the segmentation of respective lung lesions in COVID-19, several different segmentation networks have been introduced \cite{chaganti2020automated,zhang2020clinically,bai2020artificial,saood2021covid}. Whereas most of them tend to consume a lot of memory in storing the intermediate features for skip connections, it may be favorable to use several input slices in order to improve the performance in semantic segmentation tasks within data of the spatiotemporal nature \cite{cciccek20163d,milletari2016v}. We propose the application of ConvLSTM, presenting the potential to outperform other neural networks in capturing the spatiotemporal correlations, due to its capability of preserving relevant features with simultaneous dismission of irrelevant ones in the form of the feedback loop for the memory-sparing strategy and holistic analysis of the images \cite{shi2015convolutional}. It has been found that ConvLSTM localized at the input end allowed to effectively capture the global information and optimize the model performance. 

Automated segmentation of lung lesions with ConvLSTM networks offers a solution to generate big data with limited human resources and minimal hardware requirements. Since results of segmentation are presented to the human reader for visual inspection, eventual corrections enable the implementation of a human-in-the-loop strategy to reduce the annotation effort and provide high-volume training datasets to improve the performance of deep-learning models \cite{wang2020noise}. Furthermore, objective and repeatable quantification of pneumonia burden might aid the evaluation of the disease progression and assist the tomographic monitoring of different treatment responses.

Our study had several limitations. First, different patient profiles and treatment protocols between countries may have resulted in heterogeneity in COVID-19 pneumonia severity. Second, most of the CT scans were acquired during the hospital admission, therefore availability of the slices with high-opacity (consolidations and plural effusion), representing a peak stage of the disease, was limited. Finally, training and external validation datasets comprised a relatively low number of patients manually segmented by two expert readers; however, to mitigate this we have utilized repeated testing allowing us to evaluate expected average performance of the model.
\section*{Conclusion}\label{sec:con}
In conclusion, we propose and evaluate a deep learning method based on convolutional LSTM for fully automated quantification of pneumonia burden in COVID-19 patients from both non-contrast and contrast-enhanced CT datasets. The proposed method provided rapid segmentation of lung lesions with strong agreement with manual segmentation and may represent a robust tool to generate the big data with an accuracy similar to the expert readers.

\section*{Funding and acknowledgment}\label{sec:ack}
This research was supported by Cedars-Sinai COVID-19 funding. This research was also supported by the National Heart, Lung, and Blood Institute of the National Institutes of Health (R01HL133616). The content is solely the responsibility of the authors and does not necessarily represent the official views of the National Institutes of Health.
\clearpage
\section*{Figures}\label{sec:figures}
\vspace{-2em}
\newpage
\newpage
\newpage
\newpage
\newpage

\clearpage
\begin{table}[ht!]
\caption{{\bf Patient baseline characteristics and imaging data in a training cohort.}
}
\vspace{1em}
\centering
    \begin{tabular}{l@{~~}|c@{~~}}
            \toprule
            {\bf Baseline characteristics}
            & {\bf N=187} \\
            \bottomrule
            Age, years   
            & 61 ± 16 \\
            Men
            & 123 (65.7) \\
            Body mass index 
            & 26.8 ± 5.3 \\
            Current smoker
            & 22 (11.7) \\
            Former smoker
            & 10 (5.3) \\
            History of lung disease
            & 19 (10.1) \\
            \toprule
            {\bf Image characteristics}
            & {\bf N=197} \\
            \bottomrule
            CT scanner
            &  \\
            Aquilion ONE 
            & 73 (37.0) \\
            GE Revolution 
            & 13 (6.6) \\
            GE Discovery CT750 HD
            & 37 (18.8) \\
            LightSpeed VCT
            & 36 (18.3) \\
            Brilliance iCT
            & 28 (14.2) \\
            Unknown
            & 10 (5.1) \\
            Non-contrast
            & 167 (84.8) \\
            CT pulmonary angiography
            & 30 (15.2) \\
            ECG-gated
            & 35 (17.8)
    \end{tabular}
    \vspace{1em}
    \begin{tablenotes}
      \small
      \item The data presented in the table are as n(\%) or mean \textpm\xspace SD
    \end{tablenotes}
    \label{table:table1}
\end{table}
\newpage
\begin{table*}[ht!]
    \caption{{\bf Dataset information.}
    }
    \vspace{1em}
    \centering
    \begin{tabular}{c|lc|c|c|c@{~~}}\toprule
        \multicolumn{2}{c}{\multirow{2}{4em}[-1em]{Cohort}} & \multirow{2}{4em}[-1em]{No. of patients} & \multirow{2}{3em}[-1em]{No. of lesions} &
        \multicolumn{2}{c}{No. of lesion slices} \\
        \cmidrule(r){5-6}
        \multicolumn{2}{c}{} & & & 
        \multicolumn{1}{m{2cm}}{Ground-glass opacity}
        & High opacity \\
        \bottomrule
        \multicolumn{2}{c}{Training} & 197 & 31560 & 15375 & 6933 \\
        \cmidrule(r){1-6}
        \multirow{2}{5em}{External Validation} & \multicolumn{1}{m{3cm}}{Centro Cardiologico Monzino} & 17 & 10053 & 4396 & 1834 \\
        \cmidrule(r){2-6}
        & MosMedData \cite{morozov2020mosmeddata} & 50 & 2049 & 785 & 0 \\
        \bottomrule
	\end{tabular}
    \label{table:table2}
\end{table*}
\newpage
\begin{table*}[ht!]
    \caption{{\bf Computation times using different hardware components.}
    }
    \vspace{1em}
    \centering
    \begin{tabular}{c|@{~~}l|l|@{~~}}\toprule
        \multirow{2}{4em}{Hardware} & \multicolumn{2}{c}{No. of Slices} \\
        \cmidrule(r){2-3}
        & \multicolumn{1}{c}{120} & \multicolumn{1}{c}{500} \\
        \cmidrule(r){1-3}
        GPU 24GB (Titan RTX) & \multicolumn{1}{l}{1.8 secs} & \multicolumn{1}{r}{5.5 secs} \\
        GPU 4GB (Quadro K1200) & \multicolumn{1}{l}{23.5 secs} & \multicolumn{1}{r}{98 secs} \\
        CPU 3.4 GHZ (i7-6800K) & \multicolumn{1}{l}{145 secs} & \multicolumn{1}{r}{605 secs} \\
        \bottomrule
    \end{tabular}
    \label{table:table3}
\end{table*}
\newpage
\begin{table*}[ht!]
    \caption{{\bf Model Performance: Unet2D, Unet3D and Ours.}
    }
    \vspace{1em}
    \centering
    \begin{tabular}{@{}p{0.12\linewidth}|p{0.12\linewidth}|p{0.12\linewidth}|p{0.15\linewidth}|c|c|c@{}}
        \toprule
        \multicolumn{1}{c|}{\multirow{2}{4em}[-1em]{\bf Model}} & \multicolumn{3}{c|}{\bf Test Set Dice Score} & \multicolumn{3}{c}{\bf Compute}\\
        \cmidrule(lr){2-4}
        \cmidrule(lr){5-7}
        & \multicolumn{1}{c|}{\multirow{2}{*}{\bf BG}} & \multicolumn{1}{c|}{\multirow{2}{*}{\bf GGO}} & \centering{\bf High Opacities} & \multicolumn{1}{c|}{\multirow{2}{*}{\bf CPU Time}} & \multicolumn{1}{c|}{\multirow{2}{*}{\bf GPU Time}} & \multicolumn{1}{c|}{\multirow{2}{*}{\bf Memory}}\\
        \bottomrule
        \bf Unet 2D & \centering{0.99614} & \centering{\bf 0.67968} & \centering{0.55738} & 6.097s & 386.843ms & 5.80 Gb\\
        \hline
        \bf Unet 3D & \centering{0.99311} & \centering{0.61898} & \centering{0.56298} & 11.020s & 333.871ms & 18.61 Gb\\
        \hline
        \bf Ours & \centering{0.99598} & \centering{0.67944} & \centering{\bf 0.58497} & \bf 3.700s & \bf 323.421ms & \bf 4.84 Gb\\
        \bottomrule
	\end{tabular}
    \label{table:table4}
\end{table*}
\clearpage
\section*{Supplementary}\label{sec:supp}
\setcounter{figure}{0}
\renewcommand{\thefigure}{\roman{figure}}
\vspace{-2em}
\begin{figure}[ht!]
    \caption{Bland-Altman plots and Spearman correlation for volumes of ground-glass opacity (A-B) and high opacity (C-D) between expert and automatic quantification in an external validation set. 
    }
    \vspace{-2em}
    \begin{center}
        \includegraphics[width=\linewidth]{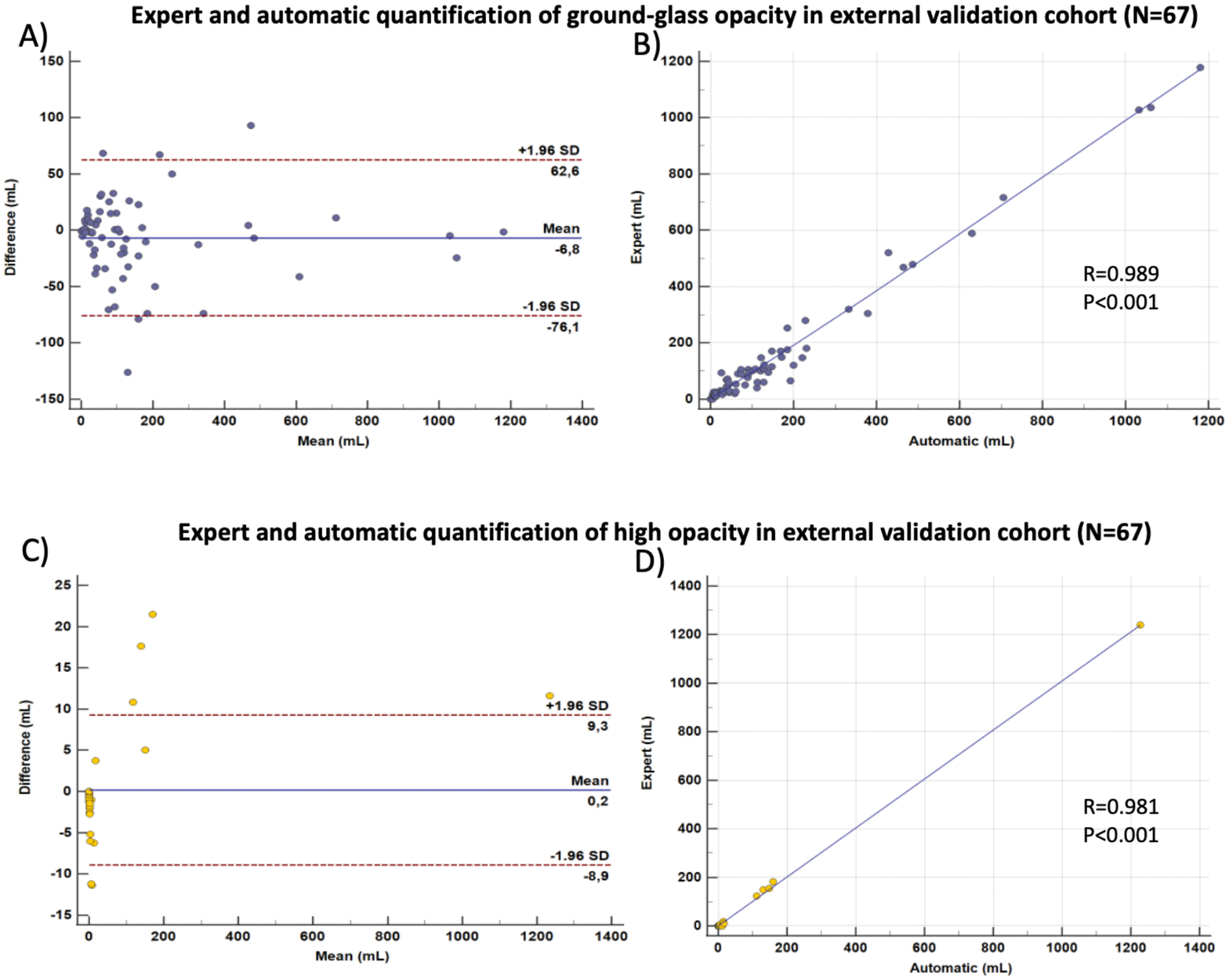}
    \end{center}
    \label{fig:supp_fig1}
\end{figure}
\newpage
\bibliographystyle{splncs04}
% \bibliographystyle{ieeetr}

% \bibliography{refs}

\begin{thebibliography}{10}
\providecommand{\url}[1]{\texttt{#1}}
\providecommand{\urlprefix}{URL }
\providecommand{\doi}[1]{https://doi.org/#1}

\bibitem{who}
{W}{H}{O} - World Health Organization Weekly epidemiological update on COVID-19 - 23 March 2021.
  \url{https://www.who.int/publications/m/item/weekly-epidemiological-update-on-covid-19---23-march-2021}

\bibitem{ai2020correlation}
Tao Ai, Zhenlu Yang, Hongyan Hou, Chenao Zhan, Chong Chen, Wenzhi Lv, Qian Tao,
  Ziyong Sun, and Liming Xia.
\newblock Correlation of chest ct and rt-pcr testing for coronavirus disease
  2019 (covid-19) in china: a report of 1014 cases.
\newblock {\em Radiology}, 296(2):E32--E40, 2020.

\bibitem{bai2020artificial}
Harrison~X Bai, Robin Wang, Zeng Xiong, Ben Hsieh, Ken Chang, Kasey Halsey, Thi
  My~Linh Tran, Ji~Whae Choi, Dong-Cui Wang, Lin-Bo Shi, et~al.
\newblock Artificial intelligence augmentation of radiologist performance in
  distinguishing covid-19 from pneumonia of other origin at chest ct.
\newblock {\em Radiology}, 296(3):E156--E165, 2020.

\bibitem{bernheim2020chest}
Adam Bernheim, Xueyan Mei, Mingqian Huang, Yang Yang, Zahi~A Fayad, Ning Zhang,
  Kaiyue Diao, Bin Lin, Xiqi Zhu, Kunwei Li, et~al.
\newblock Chest ct findings in coronavirus disease-19 (covid-19): relationship
  to duration of infection.
\newblock {\em Radiology}, page 200463, 2020.

\bibitem{boger2020systematic}
Beatriz B{\"o}ger, Mariana~M Fachi, Raquel~O Vilhena, Alexandre
  de~F{\'a}tima~Cobre, Fernanda~S Tonin, and Roberto Pontarolo.
\newblock Systematic review with meta-analysis of the accuracy of diagnostic
  tests for covid-19.
\newblock {\em American journal of infection control}, 2020.

\bibitem{chaganti2020automated}
Shikha Chaganti, Philippe Grenier, Abishek Balachandran, Guillaume Chabin,
  Stuart Cohen, Thomas Flohr, Bogdan Georgescu, Sasa Grbic, Siqi Liu,
  Fran{\c{c}}ois Mellot, et~al.
\newblock Automated quantification of ct patterns associated with covid-19 from
  chest ct.
\newblock {\em Radiology: Artificial Intelligence}, 2(4):e200048, 2020.

\bibitem{cciccek20163d}
{\"O}zg{\"u}n {\c{C}}i{\c{c}}ek, Ahmed Abdulkadir, Soeren~S Lienkamp, Thomas
  Brox, and Olaf Ronneberger.
\newblock 3d u-net: learning dense volumetric segmentation from sparse
  annotation.
\newblock In {\em International conference on medical image computing and
  computer-assisted intervention}, pages 424--432. Springer, 2016.

\bibitem{commandeur2018deep}
Frederic Commandeur, Markus Goeller, Julian Betancur, Sebastien Cadet, Mhairi
  Doris, Xi~Chen, Daniel~S Berman, Piotr~J Slomka, Balaji~K Tamarappoo, and
  Damini Dey.
\newblock Deep learning for quantification of epicardial and thoracic adipose
  tissue from non-contrast ct.
\newblock {\em IEEE transactions on medical imaging}, 37(8):1835--1846, 2018.

\bibitem{francone2020chest}
Marco Francone, Franco Iafrate, Giorgio~Maria Masci, Simona Coco, Francesco
  Cilia, Lucia Manganaro, Valeria Panebianco, Chiara Andreoli, Maria~Chiara
  Colaiacomo, Maria~Antonella Zingaropoli, et~al.
\newblock Chest ct score in covid-19 patients: correlation with disease
  severity and short-term prognosis.
\newblock {\em European radiology}, 30(12):6808--6817, 2020.

\bibitem{gieraerts2020prognostic}
Christopher Gieraerts, Anthony Dangis, Lode Janssen, Annick Demeyere, Yves
  De~Bruecker, Nele De~Brucker, Annelies van Den~Bergh, Tine Lauwerier,
  Andr{\'e} Heremans, Eric Frans, et~al.
\newblock Prognostic value and reproducibility of ai-assisted analysis of lung
  involvement in covid-19 on low-dose submillisievert chest ct: sample size
  implications for clinical trials.
\newblock {\em Radiology: Cardiothoracic Imaging}, 2(5):e200441, 2020.

\bibitem{grodecki2020quantitative}
Kajetan Grodecki, Andrew Lin, Sebastien Cadet, Priscilla~A McElhinney, Aryabod
  Razipour, Cato Chan, Barry Pressman, Peter Julien, Pal Maurovich-Horvat,
  Nicola Gaibazzi, et~al.
\newblock Quantitative burden of covid-19 pneumonia on chest ct predicts
  adverse outcomes: a post-hoc analysis of a prospective international
  registry.
\newblock {\em Radiology: Cardiothoracic Imaging}, 2(5):e200389, 2020.

\bibitem{grodecki2021epicardial}
Kajetan Grodecki, Andrew Lin, Aryabod Razipour, Sebastien Cadet, Priscilla~A
  McElhinney, Cato Chan, Barry~D Pressman, Peter Julien, Pal Maurovich-Horvat,
  Nicola Gaibazzi, et~al.
\newblock Epicardial adipose tissue is associated with extent of pneumonia and
  adverse outcomes in patients with covid-19.
\newblock {\em Metabolism}, 115:154436, 2021.

\bibitem{he2015delving}
Kaiming He, Xiangyu Zhang, Shaoqing Ren, and Jian Sun.
\newblock Delving deep into rectifiers: Surpassing human-level performance on
  imagenet classification.
\newblock In {\em Proceedings of the IEEE international conference on computer
  vision}, pages 1026--1034, 2015.

\bibitem{huang2017densely}
Gao Huang, Zhuang Liu, Laurens Van Der~Maaten, and Kilian~Q Weinberger.
\newblock Densely connected convolutional networks.
\newblock In {\em Proceedings of the IEEE conference on computer vision and
  pattern recognition}, pages 4700--4708, 2017.

\bibitem{khatami2020meta}
Fatemeh Khatami, Mohammad Saatchi, Seyed Saeed~Tamehri Zadeh, Zahra~Sadat
  Aghamir, Alireza~Namazi Shabestari, Leonardo~Oliveira Reis, and Seyed
  Mohammad~Kazem Aghamir.
\newblock A meta-analysis of accuracy and sensitivity of chest ct and rt-pcr in
  covid-19 diagnosis.
\newblock {\em Scientific reports}, 10(1):1--12, 2020.

\bibitem{krizhevsky2012imagenet}
Alex Krizhevsky, Ilya Sutskever, and Geoffrey~E Hinton.
\newblock Imagenet classification with deep convolutional neural networks.
\newblock {\em Advances in neural information processing systems},
  25:1097--1105, 2012.

\bibitem{li2020ct}
Kunwei Li, Yijie Fang, Wenjuan Li, Cunxue Pan, Peixin Qin, Yinghua Zhong,
  Xueguo Liu, Mingqian Huang, Yuting Liao, and Shaolin Li.
\newblock Ct image visual quantitative evaluation and clinical classification
  of coronavirus disease (covid-19).
\newblock {\em European radiology}, 30(8):4407--4416, 2020.

\bibitem{li2020using}
Lin Li, Lixin Qin, Zeguo Xu, Youbing Yin, Xin Wang, Bin Kong, Junjie Bai,
  Yi~Lu, Zhenghan Fang, Qi~Song, et~al.
\newblock Using artificial intelligence to detect covid-19 and
  community-acquired pneumonia based on pulmonary ct: evaluation of the
  diagnostic accuracy.
\newblock {\em Radiology}, 296(2):E65--E71, 2020.

\bibitem{milletari2016v}
Fausto Milletari, Nassir Navab, and Seyed-Ahmad Ahmadi.
\newblock V-net: Fully convolutional neural networks for volumetric medical
  image segmentation.
\newblock In {\em 2016 fourth international conference on 3D vision (3DV)},
  pages 565--571. IEEE, 2016.

\bibitem{morozov2020mosmeddata}
SP~Morozov, AE~Andreychenko, NA~Pavlov, AV~Vladzymyrskyy, NV~Ledikhova,
  VA~Gombolevskiy, Ivan~A Blokhin, PB~Gelezhe, AV~Gonchar, and V~Yu Chernina.
\newblock Mosmeddata: Chest ct scans with covid-19 related findings dataset.
\newblock {\em arXiv preprint arXiv:2005.06465}, 2020.

\bibitem{pan2020time}
Feng Pan, Tianhe Ye, Peng Sun, Shan Gui, Bo~Liang, Lingli Li, Dandan Zheng,
  Jiazheng Wang, Richard~L Hesketh, Lian Yang, et~al.
\newblock Time course of lung changes at chest ct during recovery from
  coronavirus disease 2019 (covid-19).
\newblock {\em Radiology}, 295(3):715--721, 2020.

\bibitem{saood2021covid}
Adnan Saood and Iyad Hatem.
\newblock Covid-19 lung ct image segmentation using deep learning methods:
  U-net versus segnet.
\newblock {\em BMC Medical Imaging}, 21(1):1--10, 2021.

\bibitem{shi2015convolutional}
Xingjian Shi, Zhourong Chen, Hao Wang, Dit-Yan Yeung, Wai-Kin Wong, and
  Wang-chun Woo.
\newblock Convolutional lstm network: A machine learning approach for
  precipitation nowcasting.
\newblock {\em arXiv preprint arXiv:1506.04214}, 2015.

\bibitem{tao2020hierarchical}
Andrew Tao, Karan Sapra, and Bryan Catanzaro.
\newblock Hierarchical multi-scale attention for semantic segmentation.
\newblock {\em arXiv preprint arXiv:2005.10821}, 2020.

\bibitem{wang2020noise}
Guotai Wang, Xinglong Liu, Chaoping Li, Zhiyong Xu, Jiugen Ruan, Haifeng Zhu,
  Tao Meng, Kang Li, Ning Huang, and Shaoting Zhang.
\newblock A noise-robust framework for automatic segmentation of covid-19
  pneumonia lesions from ct images.
\newblock {\em IEEE Transactions on Medical Imaging}, 39(8):2653--2663, 2020.

\bibitem{wang2020temporal}
Yuhui Wang, Chengjun Dong, Yue Hu, Chungao Li, Qianqian Ren, Xin Zhang, Heshui
  Shi, and Min Zhou.
\newblock Temporal changes of ct findings in 90 patients with covid-19
  pneumonia: a longitudinal study.
\newblock {\em Radiology}, 296(2):E55--E64, 2020.

\bibitem{xu2015empirical}
Bing Xu, Naiyan Wang, Tianqi Chen, and Mu~Li.
\newblock Empirical evaluation of rectified activations in convolutional
  network.
\newblock {\em arXiv preprint arXiv:1505.00853}, 2015.

\bibitem{yang2020chest}
Ran Yang, Xiang Li, Huan Liu, Yanling Zhen, Xianxiang Zhang, Qiuxia Xiong, Yong
  Luo, Cailiang Gao, and Wenbing Zeng.
\newblock Chest ct severity score: an imaging tool for assessing severe
  covid-19.
\newblock {\em Radiology: Cardiothoracic Imaging}, 2(2):e200047, 2020.

\bibitem{zhang2020clinically}
Kang Zhang, Xiaohong Liu, Jun Shen, Zhihuan Li, Ye~Sang, Xingwang Wu, Yunfei
  Zha, Wenhua Liang, Chengdi Wang, Ke~Wang, et~al.
\newblock Clinically applicable ai system for accurate diagnosis, quantitative
  measurements, and prognosis of covid-19 pneumonia using computed tomography.
\newblock {\em Cell}, 181(6):1423--1433, 2020.

\bibitem{zhao2019region}
Shuai Zhao, Yang Wang, Zheng Yang, and Deng Cai.
\newblock Region mutual information loss for semantic segmentation.
\newblock {\em arXiv preprint arXiv:1910.12037}, 2019.

\end{thebibliography}
% \bibliographystyle{plain}
% \input{refs.bib}
\end{document}